# Magnetic hyperthermia properties of nanoparticles inside lysosomes using kinetic Monte-Carlo simulations: influence of key parameters, of dipolar interactions and spatial variations of heating power.

R. P. Tan, J. Carrey* and M. Respaud

Université de Toulouse; INSA; UPS; LPCNO (Laboratoire de Physique et Chimie des Nano-Objets), 135 avenue de Rangueil, F-31077 Toulouse, France and CNRS; UMR 5215; LPCNO, F-31077 Toulouse, France

**Abstract:**
Understanding the influence of dipolar interactions in magnetic hyperthermia experiments is of crucial importance for fine optimization of nanoparticle (NP) heating power. In this study, we use a kinetic Monte-Carlo algorithm to calculate hysteresis loops that correctly account for both time and temperature. This algorithm is shown to correctly reproduce the high-frequency hysteresis loop of both superparamagnetic and ferromagnetic NPs without any ad-hoc or artificial parameters. The algorithm is easily parallelizable with a good speed-up behavior, which considerably decreases the calculation time on several processors and enables the study of assemblies of several thousands of NPs. The specific absorption rate (SAR) of magnetic NPs dispersed inside spherical lysosomes is studied as a function of several key parameters: volume concentration, applied magnetic field, lysosome size, NP diameter and anisotropy. The influence of these parameters is illustrated and comprehensively explained. In summary, magnetic interactions increase the coercive field, saturation field and hysteresis area of major loops. However, for small amplitude magnetic field such as those used in magnetic hyperthermia, the heating power as function of concentration can increase, decrease or display a bell shape, depending on the relationship between the applied magnetic field and the coercive/saturation fields of the NPs. The hysteresis area is found to be well correlated with the parallel or antiparallel nature of the dipolar field acting on each particle. The heating power of a given NP is strongly influenced by a local concentration involving approximately 20 neighbors. Because this local concentration strongly decreases upon approaching the surface, the heating power increases or decreases in the vicinity of the lysosome membrane. The amplitude of variation reaches more than one order of magnitude in certain conditions. This transition occurs on a thickness corresponding to approximately 1.3 times the mean distance between two neighbors. The amplitude and sign of this variation is explained. Finally, implications of these various findings are discussed in the framework of magnetic hyperthermia optimization. It is concluded that feedback on two specific points from biology experiments is required for further advancement of the optimization of magnetic NPs for magnetic hyperthermia. The present simulations will be an advantageous tool to optimize magnetic NPs heating power and interpret experimental results.

# Main Text:

## I. Introduction

Studying the properties of interacting magnetic nanoparticles (MNPs) is an old topic in magnetism, driven not only by the challenge to understand the properties of a complex many-body system but also by the will to model ferrofluids, magnetic recording media or magneto-transport properties [1, 2, 3, 4, 5, 6, 7]. More recently, a renewed interest on this topic has been motivated by the application of MNPs in magnetic hyperthermia (MH). MH properties of magnetically independent MNPs are now well understood [8, 9]. However, when MH properties of colloidal solutions are studied, MNPs aggregate under the influence of the magnetic field, which modifies their heating power compared with independent MNPs [10, 11,12]. Moreover, in *in vitro* conditions, MNPs accumulate in lysosomes, where they are highly concentrated and are thus in strong magnetic interaction [13,14]. These considerations have motivated several studies on the influence of magnetic interactions on MH properties [15, 16, 17, 18, 19, 20, 21].

Until recently, a study of the average modification of the MNP heating power due to the presence of magnetic interactions in lysosomes could have been sufficient to optimize MNPs in biological conditions. However, a series of puzzling *in vitro* results have shown that, at least in some experiments, cell death was due to a very local energy release inside the lysosomes rather than to an average global heating of the cells [13, 14, 22, 23, 24, 25, 26, 27,]. As a consequence, optimizing MNPs to maximize cell death requires an understanding of the spatial repartition of their heating properties inside lysosomes. Indeed, if cell death is, for instance, triggered by the local activation of trans-membrane receptors at the surface of the lysosomes, only the energy released by the MNPs near the lysosome surface should be maximized. In contrast, if MNPs activate a biological process occurring in the lysosome core, the heating power of the MNPs inside it should be maximized. Thus far, this spatial repartition of the heating power inside lysosomes has never been studied, and we will demonstrate that this is an important parameter to take into account.

Because the heating power of MNPs is directly given by their hysteresis loop area [8], dynamic hysteresis loop calculations using numerical simulations is the preferred method to estimate the heating power. Thus far, the main approach to perform these calculations has been Metropolis Monte-Carlo simulations [15, 17, 18]. In this algorithm, the relationship between real time and Monte-Carlo steps is not well defined [28]. This poorly defined relationship is a problem when describing MH experiments, which are conducted at a high frequency and thus require a good dynamic description. Another approach consists of using kinetic Monte-Carlo simulations, in which the dynamic is more accurately taken in consideration. Hovorka *et al.* have used this algorithm to describe magnetic recording media [29]. The advantage of this algorithm is the accurate description of MNPs in the superparamagnetic or ferromagnetic regime without any artificial or abrupt separation between them. This characteristic is important because these two classes of MNPs should be studied in MH applications: ferromagnetic MNPs display a larger specific absorption rate (SAR); however, superparamagnetic MNPs are more easily stabilized and synthesized; therefore, they are widely used *in vitro* and *in vivo*.

Here, we report the use of kinetic Monte-Carlo simulations to study the MH properties of MNPs inside lysosomes. We show the drastic influence of the volume concentration on the heating properties and present a comprehensive study on the influence of key parameters, modifying the amplitude and sign of this influence. We also show that the heating power is not homogeneous inside lysosomes and drastically depends on the position inside them.

II. Kinetic Monte-Carlo simulations
1. Algorithm

In our program, perfectly monodisperse MNPs of diameter $d$ and volume $V$, with a uniaxial anisotropy $K_{eff}$ and displaying a magnetization per unit of volume $M_S$, are considered. The main approximation of our numerical model is that we remain in the two-level approximation such that the excited states into a potential well are not taken into consideration. The MNPs have a volume concentration $c$ inside a sphere that models the lysosome. The MNPs can be either placed randomly or placed on a cubic lattice and then randomly moved a given distance to introduce disorder. A sinusoidal magnetic field of maximum amplitude $\mu_0 H_{max}$ and frequency $f$ is applied to the MNP assembly. At a given time, the total magnetic field $\overrightarrow{\mu_0 H_{tot}}$ acting on a NP is the sum of the external magnetic field $\overrightarrow{\mu_0 H_{ext}}$ and the dipolar field created by the other NPs $\overrightarrow{\mu_0 H_{dip}}$. The latter is given by:

$$\overrightarrow{\mu_0 H_{dip}} = \frac{\mu_0 M_S V}{4\pi} \sum_{i \neq j} \frac{3(\overrightarrow{m_j} \cdot \overrightarrow{e_{ij}})\overrightarrow{e_{ij}} - \overrightarrow{m_j}}{r_{ij}^3} \quad (1)$$

where $\overrightarrow{e_{ij}}$ is the unitary vector joining two NPs, $\overrightarrow{m_j}$ is the unitary vector linked to the magnetization orientation and $r_{ij}$ the distance between two NPs. This sum is calculated exactly without Ewald summation or a cutoff radius. The energy of a NP is given by:

$$E(\theta, \phi) = K_{eff} V \sin^2(\theta) - \mu_0 M_S V H_{tot} \cos(\theta - \phi), \quad (2)$$

where $\theta$ is the angle between the easy axis and the magnetization and $\phi$ is the angle between the easy axis and the total magnetic field [see Fig. 1(a)].

To calculate the hysteresis loop, time is divided into time steps $t_{step}$ during which the magnetic field is assumed constant. As a typical value, the hysteresis is divided in 2000 equal time steps. The algorithm inside the main loop of the program is the following:

i) At the beginning of a simulation step, the orientation of the magnetization in a 3D space is known for every NP because it has been calculated at the previous step.

ii) The total magnetic field acting on each particle $\overrightarrow{\mu_0 H_{tot}}$ is calculated using Equ. 1.

iii) For each NP, a 2D working plane is defined: it is the plane simultaneously containing the total magnetic field and the NP easy axis. This working plane varies at each step for each NP. The previous magnetization vector is projected onto this new plane and renormalized, permitting the calculation of the initial angle of the magnetization in this plane.

iv) The positions of the minima and maxima of the energy potentials for each NP are found. For this purpose, the $E(\theta)$ profile is discretized into 200 points. At each point, the first derivatives of the energy are calculated using the derivative of Equ. 2. If an extremum is found between two points, its precise value is determined using Newton's method. Let $\theta_1$, $\theta_2$ and $\theta_3$ be the angles of the two minima and the angle of the lower energy maximum with energies $E_1$, $E_2$ and $E_3$, respectively.

v) the initial magnetization "falls" directly into one of the two minima, following the profile of the $E(\theta)$ function. At the end of this step, one has to deal with a familiar 2D Stoner-Wohlfarth problem with thermal activation.

vi) the probability for the NP to change wells (if there are two) is calculated. The magnetization switches from the $\theta_1$ to the $\theta_2$ direction at a rate $\nu_1$ given by:

$$v_1 = v_1^0 \exp\left(-\frac{E_3 - E_1}{kT}\right). \qquad (3)$$

Similarly, the switching rate $v_2$ from the $\theta_2$ to the $\theta_1$ direction is given by:

$$v_2 = v_2^0 \exp\left(-\frac{E_3 - E_2}{kT}\right). \qquad (4)$$

In Equs. (3) and (4), $k$ is the Boltzmann constant, $T$ is the temperature, and $v_1^0 = v_2^0$ are jump attempt frequencies. If the magnetization is initially in the minimum $\theta_2$, the probability of finding it in the minimum $\theta_1$ after a time $t_{step}$ is given by [30]:

$$P(t_{step}) = \frac{v_2}{v_1 + v_2}\left(1 - \exp\left(-(v_1 + v_2)t_{step}\right)\right) \qquad (5)$$

vii) A random number between 0 and 1 is drawn. Whether the magnetization jumps depends on if the random number is greater or lower than $P(t_{step})$. In the remaining of this article, some results labeled "$T = 0$" will be shown. In these cases, the magnetization did not have the possibility to jump: this step and the previous one were skipped.

viii) return to step i).

Fig. 1(b) illustrates a few features of interest of Equs. (3)-(5): i) the relaxation time of the magnetization (equaling $\frac{1}{v_1 + v_2}$) is similar for the two minima, as illustrated by the tangents at the origins of the curves. When no external magnetic field is applied such that $v_1 = v_2$, the relaxation time of the magnetization is the well-known Néel relaxation time and is *half* of the mean time between two jumps. This point has already been discussed and illustrated in Ref. [8]. ii) For times much longer than this relaxation time, the probability of finding the magnetization in one of the two minima is independent from the initial state and tends toward $\frac{v_2}{v_1 + v_2}$; this is the superparamagnetic regime. Using the present algorithm, the magnetization of a superparamagnetic NP switches between the two minima very often and displays a very noisy hysteresis loop. However, when several hysteresis loops are averaged, this noisy curve provides the correct magnetization for the superparamagnetic NP (see below). iii) For times much shorter than the relaxation time, the magnetization remains in its initial minimum; this is the ferromagnetic regime. iv) Consequently, Equ. (5) permits the simulation of both ferromagnetic and superparamagnetic NPs without any special assumption or arbitrary separation between them. This latter point will be illustrated further below.

2. Parallelization, calculation time and typical parameters

It should be noted that steps i) to viii) can be executed in parallel because the total magnetic field acting on each particle is not calculated after each MNP magnetization movement but only once at the beginning of each step. The parallelization of the code uses an OpenMP parallelization command at the beginning of the main calculation loop to dispatch the calculation of the MNP magnetization move on several processors. The calculation time of a single hysteresis loop and the speedup due to parallelization are shown in Fig. 2. For a large number of NPs, the calculation time approximately scales with the square of the number of NPs [see Fig. 2(a)]. The results presented in this article were typically run on 32 processors and comprised 5000 MNPs with a hysteresis loop divided in 2000 steps. Using these conditions, the calculation of a single hysteresis loop lasts approximately 1 min 30 s.

Finally, most of the studies in this article were performed with the following parameters: $T = 300$ K, $K_{eff} = 13000$ Jm$^{-3}$, $M_S=10^6$ Am$^{-1}$, $\mu_0 H_{max} = 40$ mT, $f = 100$ kHz, and $v_1^0 = v_2^0 = 10^{10}$ Hz. These parameters were used in the remainder of this article when not otherwise specified. The anisotropy value corresponds to that of bulk magnetite. The magnetization value is intermediate between that of iron oxides and that of 3D magnetic metals.

### 3. Validity of the program

To assess the validity of our simulations and of the present algorithm, hysteresis loops obtained with the present program and ones obtained using the program described in Ref. [8] are compared. In Ref. [8], the basic equations describing the time evolution of the system are the same as in the present program; however, the computation of the hysteresis loop is different. In Ref [8], the *mean* magnetization of a particle is computed : the probability of locating the magnetization in the two minima is calculated, and then the corresponding mean magnetization is calculated accordingly (see Equ. (18) in Ref. [8]). Specifically, the computed hysteresis loop is the one that would be obtained if an infinite number of NPs were measured. The present program differs because the NP magnetization of a given NP is only in *one* of the two minima and is not weighted by the probability of being in one of the two. Due to this difference, computation on a large number of NPs and/or an average of a large number of hysteresis loops must be performed to obtain results similar to those obtained in Ref [8] using the present program. An additional difference is that the program in Ref [8] did not take into account magnetic interactions between NPs; therefore, the magnetic interactions were switched off in the present program, only for the purpose of comparison. The results of this comparison are shown in Figs 1(c) and 1(d) for NPs in the superparamagnetic and ferromagnetic regimes. A perfect agreement between the two programs is evidenced. Notably, a large anisotropy value was used in Fig. 1(d) ($K_{eff} = 10^6$ J/m$^3$), explaining the weak temperature dependence of the coercive field. Because the program in Ref. [8] also showed a perfect agreement with analytical results issued from the linear response theory and Stoner-Wohlfarth model-based theories, the present program is considered able to accurately calculate hysteresis loops of both superparamagnetic and ferromagnetic NPs without any special ad-hoc parameter or additional hypothesis.

### III. Results
### 1. Technical details

In this sub-section, several technical details regarding the method of placing particles, computing the hysteresis loops and averaging the data are presented. Some of these details are important to completely understand the main results of this article.

First, details on methods to address minor cycles are provided. When a typical hysteresis loop is calculated, the magnetic field switches from $+\mu_0 H_{max}$ to $-\mu_0 H_{max}$ and then back to $+\mu_0 H_{max}$. For NPs that are not perfectly saturated by $+\mu_0 H_{max}$, the hysteresis loop does not return to its initial value and is not well closed. To solve this problem, in all of our simulations, two hysteresis loops are computed one after the other. The second one is correct and is well closed. Only the second one is recorded and treated. In the remainder of this article, when it is specified that 50 hysteresis loops were run, 100 were run and 50 of them were taken into account for the data treatment.

Then, some details on the method of placing the particles inside a sphere, randomizing the hysteresis loops and averaging data are provided. To illustrate this part, a typical example in which the hysteresis area of a MNP is plotted as a function of the position inside a lysosome is shown in Fig. 3. Two different methods of placing the MNPs inside a sphere have been tested: placing them at random positions with no overlapping or placing them on a cubic lattice and then adding disorder by displacing them a random distance between 0 and a maximum disorder value. Some difference existed between the results obtained on a cubic lattice with disorder and a cubic lattice without disorder, and the disorder amplitude influenced the results (not shown). However, only minor differences existed between the results obtained with NPs placed at random positions and NPs placed on a strongly disordered cubic lattice. Experimentally, MNPs in lysosomes are strongly disordered; therefore, we have chosen to study disordered systems only. We have also arbitrarily chosen to generate the particles using the cubic lattice + disorder approach. The disorder amplitude was chosen to be as large as possible with no possible overlap between two neighboring MNPs.

In the results presented in this article, the hysteresis loops were always computed several times (typically 50) and then averaged. Two methods of obtaining average hysteresis loops were tested. The first method consists of changing the anisotropy axis direction between each loop without moving the NP position. The second method consists in changing the anisotropy axis orientation *and* the NP position between each loop. In the latter, a given NP moves between each loop around mean positions, which are the nodes of the cubic lattice. It should first be specified that, even when the hysteresis loops are averaged on several runs, the raw data extracted from any of our simulations are scattered, indicating that the heating power varies considerably from one MNP to another. A typical example of this scattering is shown in Fig. 3, in which the hysteresis area as a function of the position is plotted. Extracting interpretable results in this case requires smoothing the data with a moving average of 200 points, the result of which is also shown in Fig. 3 alongside the raw data. From this figure, it is also clear that averaging hysteresis loops on the anisotropy axis orientation and NP position leads to less noisy curves compared with averaging on the anisotropy axis orientation only. The reason for this result will be clarified later in the article (see section III.4), but can be summarized in a few words: the hysteresis area of a given particle is extremely sensitive to the exact spatial configuration of its neighbors. Therefore, varying the geometrical configuration is an efficient method to smooth the data.

2. Influence of the average volume concentration on the heating power
a) Influence of the diameter and temperature
Here, results of the average heating power of lysosomes as a function of their volume concentration $\phi$, with $\phi$ varying between $\phi = 0.01\%$ and $\phi = 30\%$, are shown. For these simulations, the number of MNPs inside the lysosome was kept constant at 3000 MNPs; therefore, the lysosome size varied in the series. The hysteresis loops were averaged on the anisotropy axis direction and NP position. Four conditions of interest were computed: i) $d = 20$ nm with a null temperature ($T = 0$), which also corresponds to the conditions obtained when studying very large diameter NPs at a finite temperature; therefore, the exact value of the diameter does not matter for this condition. In the remainder of the article and in the figure legends, this condition will be referred to as the $d \to \infty$ case. ii) $d = 20$ nm, $T = 300$ K. iii) $d = 9$ nm, $T = 300$ K. iv) 9 nm NPs with a null anisotropy ($K_{\text{eff}} = 0$) at $T = 0$. Again, identical results are obtained for NPs with different diameters; therefore, the exact value of the diameter does not matter and will be omitted in the legends and in the discussion.

The hysteresis area as a function of $\phi$ for these four conditions is shown in Fig. 4(a), and corresponding hysteresis loops are shown in Fig. 5. In Fig. 4(a), the hysteresis area is given in J/m$^3$. To calculate the corresponding SAR, the density $\rho$ of the material (for magnetite, $\rho \approx 5200$ kg/m$^3$) and frequency should be taken into account. For instance, 10000 J/m$^3$ corresponds to a SAR of 192 W/g.

The case of $\phi = 0.01\%$, when the NPs are almost magnetically independent, is first described. The $d \to \infty$ result confirms that the magnetic parameters of the MNPs ($K_{eff}$ and $M_S$) are well adapted to the magnetic field amplitude, i.e., the hysteresis loop is well opened and saturated by the applied magnetic field [see Fig. 5(a)], leading to a large hysteresis area [see Fig. 4(a)]. As shown in Ref [8], ideal NPs for magnetic hyperthermia are large, single-domain NPs with a low anisotropy, and the hysteresis curve for $\phi = 0.01\%$ displayed in Fig. 5(a) is typical for these NPs. NPs 20 nm in size have a reduced coercive field and heating power compared with the $d \to \infty$ case due to the finite temperature and diameter [see Figs. 4(a) and 5(b)]. However, their heating power remains rather large. In contrast, NPs 9 nm in size are clearly superparamagnetic in these conditions and display a negligibly small heating power [see Figs. 4(a) and 5(c)].

Next, the effect of increasing $\phi$ on the heating power of 9 nm and 20 nm NPs is described. Figs 5(b) and 5(c) show that the effect on the hysteresis loop is rather similar in the two cases: the magnetic interactions make the saturation of the NPs more difficult; therefore, the magnetization value at 40 mT shows a monotonic decrease with increasing $\phi$. Simultaneously, the coercive field first increases with $\phi$ and then decreases. The global effect is an increase in the heating power followed by a decrease. The maximum heating power occurs in the range of 0.6-2% for $\phi$. Interestingly, the 9 nm NPs, which were not heated at all at low concentrations, show a very large heating power for $\phi = 1\%$, which is similar to the 20 nm NPs. In the $d \to \infty$ case, the behavior is similar except for heating power increases very weakly at small concentrations (only a few percent). A deeper insight on the origin of these behaviors will be provided in section III.2.b.

For $\phi$ values larger than approximately 3%, the hysteresis loops all converge toward a common shape, which is independent of the NP volume and is the same as NPs without any anisotropy [see Figs. 4(a) and 5(a)-(d)]. This regime is characterized by the fact that the NP properties are completely dominated by magnetic interactions; therefore, it is termed the "dipolar regime". In Figs. 4(a) and 5(d), the hysteresis loops and hysteresis area value of this dipolar regime ($K_{eff} = T = 0$) as a function of $\phi$ are shown. This regime is characterized by a moderate heating power due to a rather small coercive field, and has a maximal heating power a $\phi$ equal to approximately 1%.

Interestingly, a magnetic parameter that correlates well with the increase or decrease of SAR in the lysosome has been found. When the hysteresis loops were computed, the evolution of the dipolar field acting on each particle, projected in the direction of the external magnetic field, was also computed and termed $\mu_0 H_{dip}^x(\mu_0 H)$. These computations were performed to investigate if the dipolar field was locally increasing or decreasing the external magnetic field. From each dipolar field hysteresis loop, the following parameter was extracted:

$$\Delta^-_{Hdip} = \mu_0 H_{dip}^x(+\mu_0 H_{max}) - \mu_0 H_{dip}^x(-\mu_0 H_{max}) \qquad (6)$$

where $\Delta^-_{Hdip}$ is positive (negative) if the projected dipolar field is parallel (antiparallel) to the external magnetic field. A strong correlation between the amplitude/sign of $\Delta^-_{Hdip}$ and the effect of the dipolar interactions on the $A$ value has been found. To illustrate this point, Fig. 4(b) shows the evolution of the $\Delta^-_{Hdip}$ value averaged for all of the lysosomes and for the same parameters as those used in Fig. 4(a). Comparing Figs. 4(a) and 4(b) illustrates the similarity between the evolution of $A$ and $\Delta^-_{Hdip}$: the ferromagnetic (antiferromagnetic) nature of the magnetic interactions correlates with an increase (decrease) in area. This correlation is also found at the level of individual particles inside a lysosome. To illustrate this point, in Fig. 4(c), the value of $\Delta^-_{Hdip}$ as a function of $A$ for each particle inside a lysosome is plotted. The example shown corresponds to $d = 20$ nm, $\phi = 3\%$ and $T = 300$ K. A similar correlation between a positive value of $\Delta^-_{Hdip}$ and a large hysteresis area has been found in all of the other cases.

b) Influence of the applied magnetic field.
The influence of the magnetic field amplitude on the previous results is now presented. In the simulations presented above, the applied magnetic field was large enough to saturate the hysteresis loops of independent NPs, but not those of interacting MNPs. The influence of the concentration in a case where $\mu_0 H_{max} = 1$ T, with the other parameters being the same as previously described, is first presented. The magnetic field amplitude is clearly not reasonable for magnetic hyperthermia; however, the results obtained with this value are instructive. The evolution of the hysteresis area as a function of $\phi$ is shown in Fig. 6(a), and the corresponding hysteresis loops are shown in Fig. 6(c). Applying 1 T saturates the hysteresis loops even for large concentrations. The influence of magnetic interactions on the hysteresis loop shape and area is rather simple: the coercive field, saturation field and hysteresis area all increase monotonously as a function of the concentration.

Second, the results for a much lower value of $\mu_0 H_{max}$ are presented. The magnetic field value chosen for this example is the one for which 20 nm NPs are perfectly optimized. To maximize the hysteresis loop area in a given magnetic field, the following relationship between the applied magnetic field and the MNP coercive field should be verified (see Equ. 42 in Ref. 8):

$$\mu_0 H_{max} \approx \frac{\mu_0 H_C}{0.81} \qquad (7)$$

Simulations using $\mu_0 H_{max} = 7$ mT have been performed to verify Equ. (7). This condition corresponds to a practical case where, for a given imposed external magnetic field of 7 mT and a given imposed material with $K_{eff} = 13000$ J/m$^3$ and $M_S = 10^6$ Am$^{-1}$, optimizing the heating power by varying the NP size leads to a diameter of 20 nm. For these imposed conditions, 20 nm NPs are perfect NPs with the hypothesis that magnetic interactions are negligible. In Fig. 6(a), the evolution of their heating power with concentration is shown. The corresponding hysteresis loops are shown in Fig. 6(b). For these optimized NPs, magnetic interactions have catastrophic consequences because their heating power is almost completely canceled for $\phi$ values as small as 0.6%. Indeed, these NPs have a coercive field very close to the applied magnetic field when they are independent; therefore, they cannot be switched by the external magnetic field after increasing their coercive field due to magnetic interactions.

All of the results shown in Fig. 6 can be easily and qualitatively understood. Magnetic interactions increase the coercive and saturation fields of the major loops. When the applied magnetic field is larger (smaller) than these fields, the SAR increases (decreases) with the interactions. This result explains well that a monotonous increase, a monotonous decrease, or a bell shape curve is observed, depending of the relationship between the applied magnetic field and the coercive/saturation field of the major loops.

c. Influence of the anisotropy, magnetization, and universal curves.

The influence of magnetic interactions has been shown above for examples where $K_{eff}$ and $M_S$ were held constant. These two parameters are expected to have a drastic influence on the final result. Saturation magnetization enhances the effect of the magnetic interactions, whereas anisotropy should decrease their influence [7, 18]. Simulations were run at $T = 0$ with varying magnetization and anisotropy. Again, this condition also corresponds to large NPs ($d \rightarrow \infty$). The hysteresis area as a function of $\phi$ was calculated. Because increasing $K_{eff}$ increases the magnetic field required to saturate the MNPs, the magnetic field was adapted in each simulation to obtain only the major hysteresis loops *at low concentrations*. Specifically, the ratio $\frac{\mu_0 H_{max}}{K_{eff}}$ was kept constant in the series. A first series of simulations with $M_S = 10^6$ Am$^{-1}$ and $K_{eff}$ in the range of $6 \times 10^3$-$1 \times 10^6$ Jm$^{-3}$, as well as a series with $K_{eff} = 13 \times 10^3$ Jm$^{-3}$ and $M_S$ in the range of $2 \times 10^5$-$2 \times 10^6$ Jm$^{-3}$ were run. The results are shown in Fig. 7 as black curves. All of the obtained data converge to a single curve in a plot of a dimensionless hysteresis area $\frac{A}{K_{eff}}$ as a function of a dimensionless concentration $\frac{\mu_0 M_S^2 \phi}{K_{eff}}$. The dimensionless area equals 2 at low concentrations, which is the hysteresis loop area of randomly oriented NPs (see Equ. 14 in Ref [8]). At large concentrations, the hysteresis area decreases because the applied magnetic field is below the coercive and/or saturation field of the MNPs. Therefore, this curve is a generalization of the $d \rightarrow \infty$ curve shown in Fig. 4(a). The blue curves in the same figure represent the influence of $\mu_0 H_{max}$ for a constant $K_{eff}$. The behavior is the same as that observed in Fig. 6(a). The red curves represent the influence of a finite diameter for a constant $K_{eff}$ and a constant $\mu_0 H_{max}$. Therefore, a correspondence exists between these red curves and the study on the influence of diameter shown in Fig. 4(a). Two findings can be extracted from this figure: i) large NPs follow a series of universal curves (the blue and black lines) in which the dimensionless area only depends on the dimensionless concentration. Each universal curve corresponds to a given $\frac{\mu_0 H_{max}}{K_{eff}}$ ratio. As previously discussed, these universal curves decrease, increase or have a bell shape. ii) When the normalized concentration is below 0.02, there is no influence of the magnetic interactions, even at a finite temperature or diameter.

4. Spatial dependence of the heating power inside lysosomes and the influence of the number of NPs.

The previous section presented results averaged on all of the NPs inside the lysosome. Now, an investigation of the heating power variation at different locations inside the lysosome is presented. Fig. 8 displays the evolution of the heating power as a function of the normalized distance from the lysosome center for 20 and 9 nm NPs and for $\phi$ ranging from 0.01% to 30%. These parameters are the same as those used in previous sections; therefore, there is a strong relationship between Fig. 4(a) and Fig. 8. Fig. 8 shows the spatial dependence of the average heating power displayed in Fig. 4(a). The heating power inside the lysosome cores is constant or displays a low amplitude and smooth variation; however, it may vary considerably near their surfaces: inside some of them, the heating power varies abruptly and strongly when the normalized distance is in the range of 0.86-1. This result means that this transition occurs at a thickness corresponding to approximately 1.3 times the mean distance between two neighboring NPs. Depending on the $\phi$ value, this variation can be an increase or a decrease, as well as having a different amplitude. For instance, a lysosome filled at 0.6% with 9 nm NPs displays a heating power 14 times smaller near its surface than in its center, whereas, if it is filled at 3%, heating power is 6 times larger at the surface.

All of the curves in Fig. 8 can be qualitatively understood by considering that NPs near the lysosome surface have fewer neighbors and a lower effective concentration. When considering the curves in Fig. 4(a), this result means that approaching the surface is equivalent to a displacement to the left of these curves. Thus, the sign and amplitude of the derivative of these curves precisely explain the behavior observed in Fig. 8. For instance, the two strong variations given as examples in the previous paragraph correspond to two points that have large derivative values and opposite signs in Fig. 4(a). This explanation is provided to the reader as a first approach of the underlying mechanism. In next section, this preliminary explanation will be completed.

To visually illustrate these spatial variations of heating power, lysosomes loaded with NPs are shown in Fig. 9 with a color map corresponding to the NP heating power. Contrary to the graphs shown in Fig. 8, this figure has no spatial averaging of the heating power; therefore, the scattering of the heating power inside the lysosomes is more visible than in Fig. 8. Despite this difference, the variation of heating power between the center and the surface of the lysosome is clearly observed.

Finally, in Fig. 10, the evolution of the heating power with concentration is plotted for lysosomes filled with a number of 9 nm NPs varying between 10 and 5000. In all cases, the heating power as a function of the concentration curves presents a bell shape similar to Fig. 4(a). However, the curves are shifted: to display a heating power similar of the one of the 5000 NP lysosomes, a 100 NP lysosome must be more concentrated. Again, this result can be qualitatively understand using concepts that have been used previously to explain the spatial variation of heating power: compared to a lysosome with many NPs, a lysosome with a low number of NPs has a larger surface area / volume ratio and comparatively more particles displaying a reduced effective concentration at its surface. Its average concentration must be increased to display behavior similar to a larger lysosome.

5. Importance of the local concentration

The simulation results presented in the previous section evidence the strong influence of the local environment of a given NP on its heating power. We hypothesized that the local volume concentration around a given NP might be one of the main parameter governing its heating power because surface effects in the lysosome seem to be confined to an extremely thin layer. Different local concentrations can be defined depending on the number of neighbors taken into account. The quantity $\phi_N$, which is the local volume concentration around a given particle taking into account $N$ neighbors, is introduced. It is defined as:

$$\phi_N = \frac{3NV}{4\pi r_N^3} \qquad (8)$$

where $r_N$ is the distance between the $N^{th}$ neighbor and the particle under consideration and $V$ the NP volume.

This objective of this study is to investigate if a single curve could be obtained by only plotting the heating power of NPs as a function of the local concentration, independent of the lysosome mean concentration. For this purpose, we ran simulations on lysosomes containing 9 and 20 nm with the concentration varying between 0.01% and 30%. The parameters were the same as those used to plot Fig. 4(a) except that, in the present case, averaging was performed on the anisotropy axis orientation but not on the position. Otherwise, the calculated local concentration around a given NP would have been averaged and been rendered meaningless. We then plotted the obtained heating power as a function of various values of the local concentration ($\phi_5$, $\phi_{10}$, $\phi_{20}$, $\phi_{50}$, $\phi_{100}$, and $\phi_{200}$) to determine which value led to the best universal plotting. The best result, obtained using $\phi_{20}$, is shown in Fig. 11. For a given particle diameter, all of the data collapse well onto a single and smooth curve. These two curves are more fundamental than those shown in Fig. 4. In the latter, the volume concentration and hysteresis area varied from point to point, especially when approaching the surface. Therefore, the curves in Fig. 4 are a convolution of the curves in Fig. 11. Notably, $\phi_{20}$ corresponds to a local concentration inside a sphere of radius equaling approximately 1.6 times the mean distance between neighbors. The thickness near the lysosome membrane in which the transition of heating power occurs was 1.3 times the mean distance between neighbors (see section III.4). These two different approaches converge to a similar result.

IV. Discussion

We first summarize and provide a global view of the results obtained in the results section, starting with the influence of magnetic interactions on the heating power. First, the influence of magnetic interactions on the hysteresis loops and heating power become noticeable only when $\frac{\mu_0 M_s^2 \phi}{K_{eff}} > 0.02$. Above this value, magnetic interactions have two effects on the major hysteresis loops: i) increase the coercive field and ii) incline the hysteresis loop and increase the saturation field. The increase of coercivity of such NPs for a moderate amount of magnetic interactions has been previously shown using metropolis Monte-Carlo simulations [5, 18] and the Fokker-Planck equation [21] but not using LLG [20].

The consequence of the coercive field increase is that, for a very large applied magnetic field, the heating power monotonously increases with concentration (see the curves at large fields in Fig. 6(a) and 7). However, using the magnetic field values currently used in magnetic hyperthermia, there is a concentration above which the applied magnetic field is too small to saturate or rotate the particles, leading to a decline of the heating power when increasing the concentration and resulting in a bell shaped SAR ($\phi$) curves. Finally, if the applied magnetic field is below the coercive or saturation field of the MNPs, magnetic interactions monotonously decrease the SAR, even at low concentrations. In summary, the SAR ($\phi$) curve can decrease, increase or have a bell shape depending on the relationship between the applied magnetic field and the coercive/saturation field of the major loops.

These findings have important practical consequences for magnetic hyperthermia optimization. Two types of MNPs for magnetic hyperthermia can be defined, and the generalities presented above permit an understanding of their behavior in standard hyperthermia conditions: i) low-anisotropy ferromagnetic MNPs, once optimized, can display very large SAR values when they are magnetically independent [8]. At low concentrations, their SAR increases or decreases, but maintains the same order of magnitude (see Fig. 4(a) and 7). ii) Superparamagnetic MNPs display a much lower SAR at low concentrations. However, they are easier to synthesize, stabilize, handle and make stealthy than ferromagnetic MNPs. Once concentrated, their heating power can increase by several orders of magnitude, and their SAR can reach values similar to those of ferromagnetic NPs in the regime [see Fig. 4(a)]. The strategy of using concentration to increase the SAR of superparamagnetic NPs holds only at the condition that the dispersion of the local NP concentration inside the cells is not too large because their SAR strongly varies with the local concentration.

For a large degree of interactions, all types of NPs have strongly reduced SARs, except for those with very large anisotropies. Therefore, one strategy could consist of using high anisotropy NPs and large magnetic fields to simultaneously obtain SAR values and NPs insensitive to magnetic interactions. To quantify this approach, let us imagine that one wants to use NPs insensitive to magnetic interactions for a local concentration of $\phi = 30\%$ with $M_S = 10^6$ Am$^{-1}$. Strictly verifying the condition $\frac{\mu_0 M_S^2 \phi}{K_{eff}} < 0.02$ would require NPs with an anisotropy of approximately $1.8 \times 10^6$ J/m$^3$ (above that of Co) and a working magnetic field in the Tesla range to saturate them. These conditions are completely different from the ones currently used in magnetic hyperthermia where, due to the small amplitude of the magnetic field, low anisotropy MNPs must be used [8]. Such large amplitude magnetic fields at a reasonably high frequency, (approximately 2 kHz) to maintain a constant $\mu_0 H_{max} f$ product and thus similar SAR values than at 100 kHz, 20 mT, are so far technically inaccessible. However, it appears that increasing the magnetic field amplitude, decreasing its frequency, and increasing the MNP anisotropy is the only way to combine insensitivity to magnetic interactions and large SAR values

In the absence of such conditions, i.e., with the current values of applied magnetic field, we must address the concentration dependence of SAR. The only solution is to calculate the optimal characteristics of MNPs once the local concentrations from biology experiments are known. At present, such data are still lacking. However, general principles on the mechanisms involved in this optimization have been presented in this article.

Figs. 8 and 9 show that heating power inside the lysosomes strongly varies depending on the position inside them. When cell death or tumor regression is due to a global increase of cells or tumor temperature, this variation has no important consequences since only the average heating power is optimized. However, it has been shown that in many cases, cell death in *in vitro* experiments is not due to a global temperature increase [13, 14, 22, 23, 24, 25, 26, 27] and involves lysosome membrane permeabilization [13, 14]. This effect has not yet been explained; however, one can reasonably hypothesize that a phenomenon triggered by a local energy release occurs in the immediate vicinity of the NPs. Then, there are two possibilities: i) this phenomenon directly damages the lysosome membrane so that only particles near it would contribute to the effect and ii) the membrane permeabilization is an indirect consequence of a phenomenon occurring in the core of the lysosome. Depending on which of these two hypotheses is true, the NPs for optimizing cell death are not the same. For the former (latter) case, the heating power of the NPs at the surface (in the core) should be maximized. It has been shown that the difference between the heating power at the surface or in the core can reach one order of magnitude; therefore, the discovery of the microscopic origin of cell death in these puzzling *in vitro* experiments will permit further improvements of NPs to maximize the cell damage.

V. Conclusions.

We have studied the influence of magnetic interactions on magnetic hyperthermia properties and found that local concentration considerably affects the heating power amplitude. The increase or decrease of the hysteresis area is well correlated to the parallel or anti-parallel nature of the projected dipolar field acting on each particle. One central parameter influencing the SAR value is the volume concentration of the 20 nearest neighbors around a given NP. This sensitivity to the local concentration leads to a spatial variation of heating power as a function of the position inside the lysosome, especially near the membrane where the SAR variation compared with the core can be very large. The influence of magnetic interactions strongly depends on NP diameter and anisotropy as well as on the amplitude of the applied magnetic field. These effects can be summarized as follows: i) as expected, increasing the anisotropy decreases the effect of interactions. ii) The NP diameter is of crucial importance because it drags the NPs from a superparamagnetic regime to a ferromagnetic regime, in which the sensitivity to magnetic interactions and the maximum SAR are very different. iii) Magnetic interactions increase the coercive field, saturation field and hysteresis area of the major loops. However, in the minor loops, depending on the applied magnetic field value and its relationship with the MNP coercive field, a decrease, an increase or a non-monotonic variation of the SAR with concentration might be observed.

Because the local concentration of NPs might vary in cells or tumors, decreasing the influence of magnetic interactions may result in a constant and large SAR value, independent of the environment. The only way to achieve this result would be to use NPs with large anisotropy. However, this simplistic solution faces is limited by the maximum applied magnetic field in hyperthermia experiments. The optimal characteristics of the NPs result from a compromise where the entrance parameters are i) the maximum applied magnetic field, ii) the maximum acceptable diameter, iii) the local concentration of NPs inside the cells or tumors, and iv) the location inside the lysosomes where the heating power should be maximized. When this information is available, the numerical simulations that we have developed will be an advantageous tool to predict the heating power of MNPs in *in vivo* or *in vitro* conditions and to calculate the anisotropy required to maximize the heating power.

**Acknowledgements:**
This work was supported by the ITMO Cancer during the "Plan cancer 2009-2013" and was performed using HPC resources from CALMIP (Grant 2014-P1405). We thank Nicolas Renon for his help with the code parallelization.

**References:**
* Electronic mail: julian.carrey@insa-toulouse.fr

[1] D. Kechrakos and K.N. Trohidou, Phys. Rev. B **58**, 12169 (1998).
[2] D. Kechrakos and K.N. Trohidou, Appl. Surf. Sci. **226**, 261(2004)
[3] C. Verdes, B. Ruiz-Diaz, S. M. Thompson, R. W. Chantrell and Al. Stancu, Phys. Rev. B **65**, 174417 (2002).
[4] S. Russ and A. Bunde, Phys. Rev. B **75**, 174445(2007)
[5] N. L. Tran and H. H. Tran, J. Non. Crys. Sol. **357**, 996 (2011)
[6] M. Rekveldt and R. Rosman, J. Magn. Magn. Mater. **95**, 221 (1991)
[7] R. P. Tan, J. S. Lee, J. U. Cho, S. J. Noh, D. K. Kim and Y. K. Kim, J. Phys. D: Appl. Phys. **43**, 165002 (2010)
[8] J. Carrey, B. Mehdaoui and M. Respaud, J. Appl. Phys. **109**, 033901 (2011)
[9] N. A. Usov, J. Appl. Phys. **107**, 123909 (2010)

[10] B. Mehdaoui, R. P. Tan, A. Meffre, J. Carrey, S. Lachaize, B. Chaudret and M. Respaud Phys. Rev. B **87**, 174419 (2013)
[11] B. Mehdaoui, A. Meffre, L.-M. Lacroix, J. Carrey, S. Lachaize, M. Goujeon, M. Respaud, and B. Chaudret, J. Magn. Magn. Mat. **322**, L49 (2010)
[12] S. L. Saville, B. Qi, J. Baker, R. Stone, R. E. Camley, K. L. Livesey, L. Ye, T. M. Crawford and O. T. Mefford, J. Colloid Interface Sci. **424**, 141–151 (2014)
[13] C. Sanchez, D. El Hajj Diab, V. Connord, P. Clerc, E. Meunier, B. Pipy, B. Payré, J. Carrey, V. Gigoux and D. Fourmy, ACS Nano **8**, 1350 (2014)
[14] M. Domenech, I. Marrero-Berrios, M. Torres-Lugo and C. Rinaldi, ACS Nano **7**, 5091 (2013)
[15] V. Singh and V. Banerjee, J. Appl. Phys. **112**, 114912 (2012)
[16] F. Burrows, C. Parker, R. F. L. Evans, Y. Hancock, O. Hovorka and R. W. Chantrell, J. Phys. D : Appl. Phys. **43**, 474010 (2010)
[17] D. Serantes, D. Baldomir, C. Martinez-Boubeta, K. Simeonidis, M. Angelakeris, E. Natividad, M. Castro, A. Mediano, D.-X. Chen, A. Sanchez, Ll. Balcells, and B. Martínez, J. Appl. Phys. **108**, 073918 (2010)
[18] C. Martinez-Boubeta, K. Simeonidis, D. Serantes, I. Conde-Leborán, I. Kazakis, G. Stefanou, L. Peña, R. Galceran, L. Balcells, C. Monty, D. Baldomir, M. Mitrakas and M. Angelakeris, Adv. Func. Mater. **22**, 3737 (2012).
[19] C. L. Dennis, A. J. Jackson, J. A. Borchers, P. J. Hoopes, R. Strawbridge, A. R. Foreman, J. van Lierop, G. Grüttner and R. Ivkov, Nanotechnology **20**, 395103 (2009)
[20] C. Haase and U. Nowak, Phys. Rev. B **85**, 045435 (2012)
[21] G. T. Landi, Phys. Rev. B **89**, 014403 (2014)
[22] M. Creixell, A. C. Bohorquez, M. Torres-Lugo and C. Rinaldi, ACS Nano **5**, 7124 (2011)
[23] I. Marcos-Campos, L. Asin, T. E. Torres, C. Marquina, A. Tres, M. R. Ibarra and G. F. Goya, Nanotechnology **22**, 205101 (2011)
[24] V. Grazú, A. M. Silber, M. Moros, L. Asín, T. E. Torres, C. Marquina, M. R. Ibarra and G. F. Goya, Int. J. Nanomed. **7**, 5351 (2012)
[25] L. Asín, M. R. Ibarra, A. Tres and G. F. Goya, Pharm. Res. **29**, 1319 (2012)
[26] A. Villanueva, P. de la Presa, J. M. Alonso, T. Rueda, A. Martinez, P. Crespo, M. P. Morales, M. A. Gonzalez-Fernandez, J. Valdes and G. Rivero, J. Phys. Chem. C **114**, 1976 (2010)
[27] L Asín, G. F. Goya, A Tres and M. R. Ibarra, Cell Death and Disease **4**, e596 (2013)
[28] P. V. Melenev, Yu. L. Raikher, V. V. Rusakov and R. Perzynski, Phys. Rev. B. **86**, 104423 (2012)
[29] O. Hovorka, R. F. L. Evans, R. W. Chantrell and A. Berger, Appl. Phys. Lett. **97**, 062504 (2010)
[30] R. W. Chantrell, N. Walmsley, J. Gore and M. Maylin, Phys. Rev. B **63**, 024410 (2000)

**Figures:**

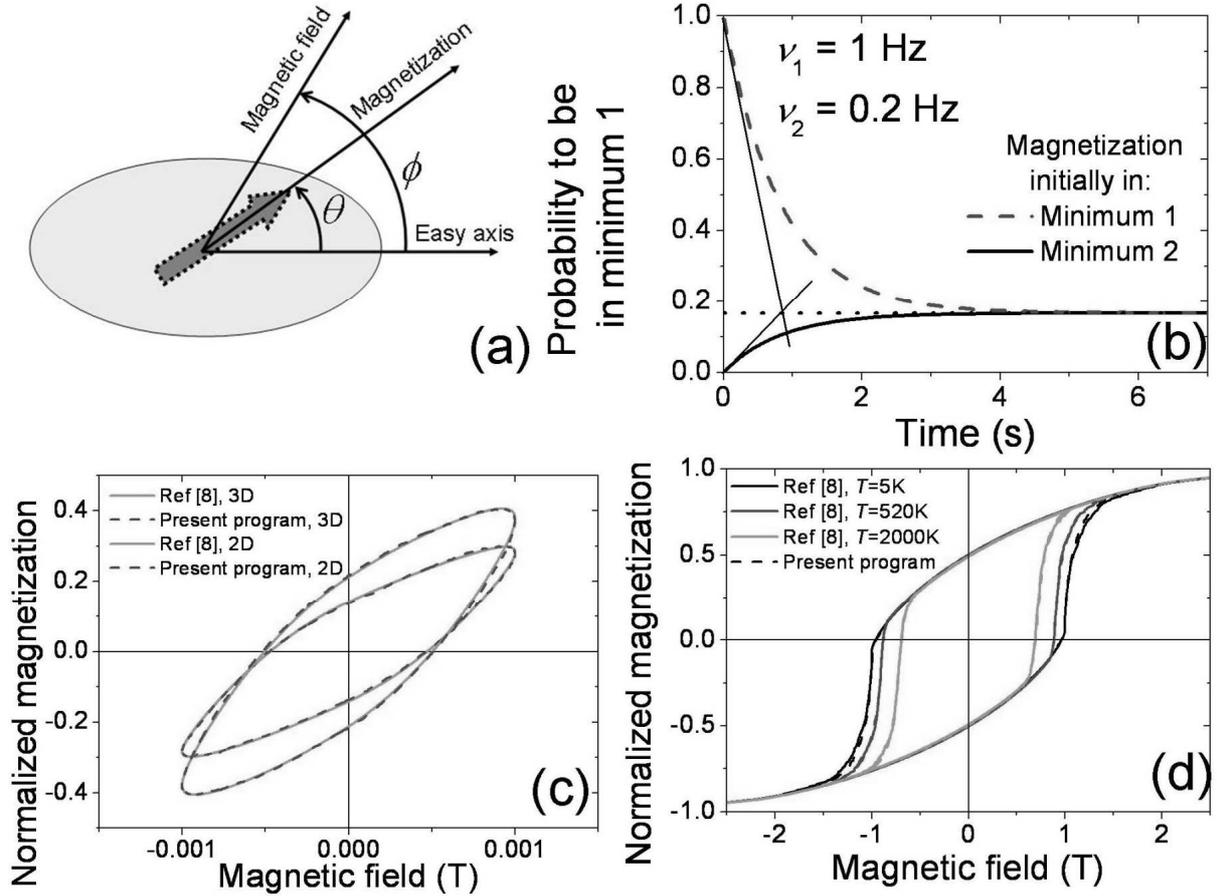

Figure 1 (color online): (a) Schematic of a NP illustrating the angles used in the main text. (b) Illustration of the algorithm used in the simulations. The probability of finding the magnetization into the first minimum is plotted. It is assumed that, at $t = 0$, the magnetization is in the first (dashed line) or second (plain line) minimum. The jumping rate from the first to the second minimum $\nu_1 = 1$ Hz. The reciprocal jump rate $\nu_2 = 0.2$ Hz. The horizontal dotted line represents the probability at infinite time, here equaling 0.166. The two thin plain lines represent the initial slopes of the curves. (c) and (d) Comparison between the numerical calculation in Ref. [8] and the present program. For this comparison, magnetic interactions have been switched off in the present program. Magnetic field and NP parameters have been chosen in order to obtain hysteresis loops typical of (c) the superparamagnetic regime and (d) the ferromagnetic regime. In all cases, the hysteresis loops obtained by the two programs are almost perfectly superimposed. (c) Results from a case where the NP anisotropy axes are randomly oriented in space (labeled 3D) and randomly oriented in a 2D plane containing the applied magnetic field (labeled 2D). For the present program, 6000 point hysteresis loops were run 10 times in the raw on 4000 NPs (2D case) or 2000 NPs (3D case) and then averaged. $K_{\text{eff}} = 10^4$ J/m$^3$, $d = 20$ nm, $\mu_0 H_{\text{max}} = 1$ mT. (d) Three hysteresis loops performed at $T = 5$, 520 and 2000 K are shown. The NP anisotropy axes were randomly oriented in space. For the present program, 10000 step hysteresis loops were run on 400 NPs. $K_{\text{eff}} = 10^6$ J/m$^3$, $d = 12$ nm, $\mu_0 H_{\text{max}} = 2.5$ T.

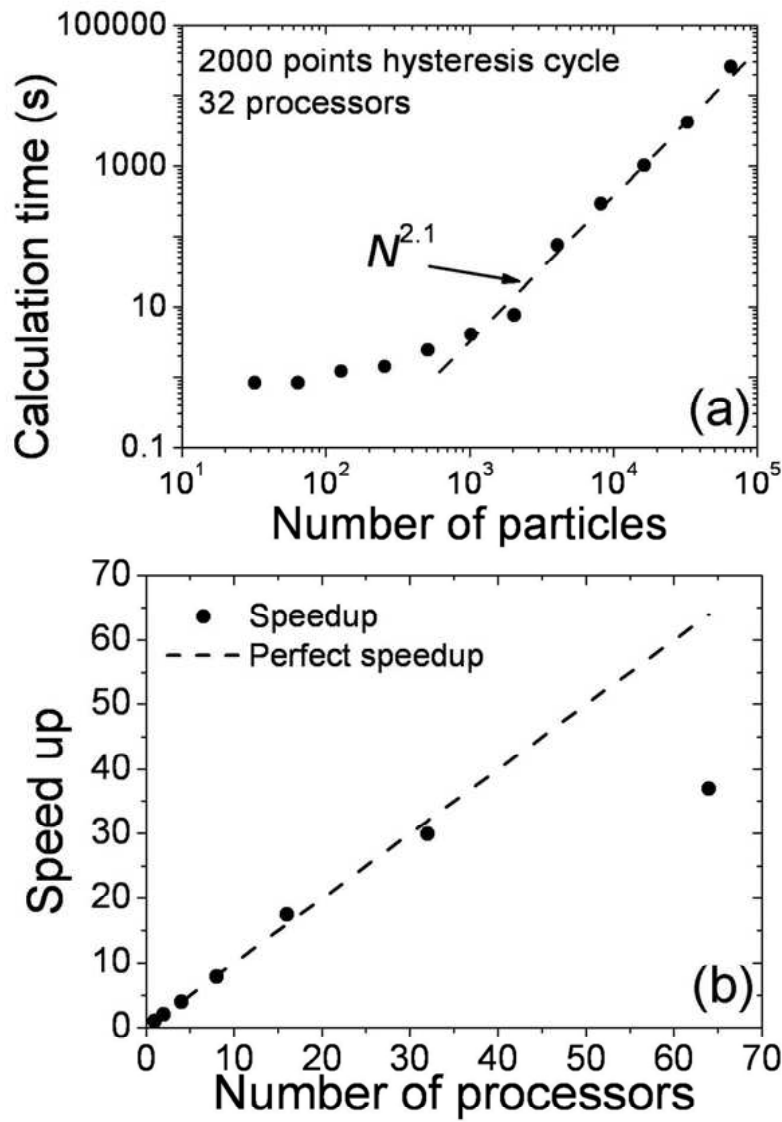

Figure 2: (a) Calculation time of a single hysteresis loop as a function of the number of NPs. Each hysteresis loop was composed of 2000 steps. The program was run on 32 processors in parallel. For a large number of NPs, a power exponent of 2.1 is found, as illustrated by the dashed line. (b) Speed-up as a function of the number of processors on which the program is run. A 2500 step hysteresis loop of 1000 NPs was computed. The dashed line represents the perfect speed-up, equaling the number of processors.

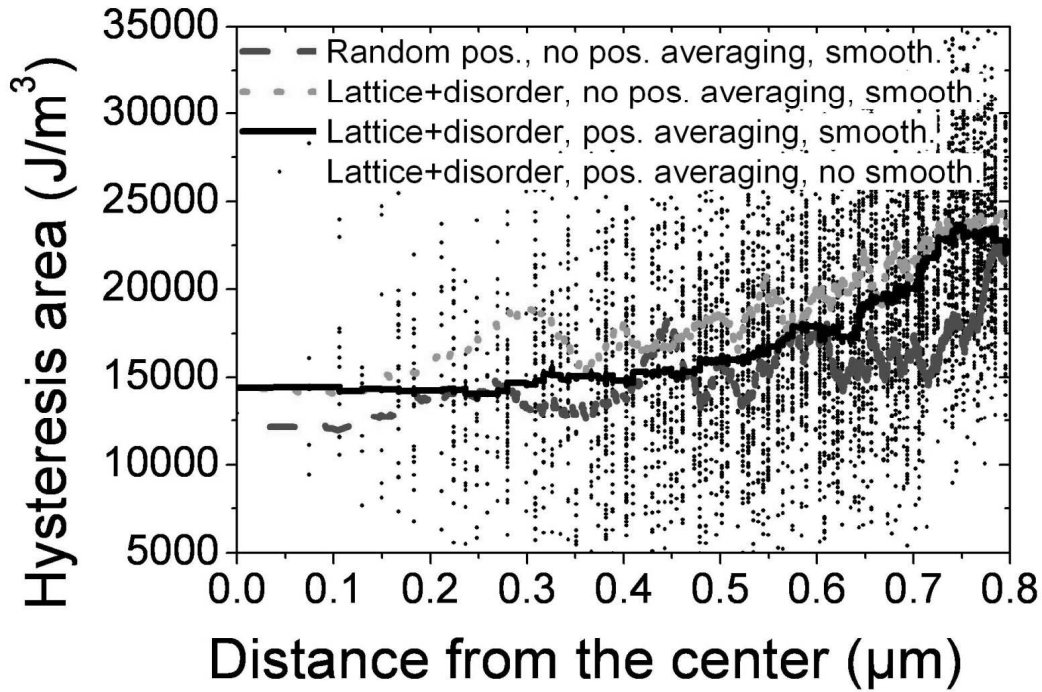

Figure 3 (color online): Illustration of the different methods of averaging. The example shown is a calculation of the hysteresis area as a function of the position inside a 5000 NP lysosome, with $d = 20$ nm, $c = 0.01$, and $\mu_0 H_{max} = 40$ mT. For all of the data shown, 50 hysteresis loops were calculated and averaged. The dashed line represents a calculation in which the NPs were randomly placed. For the other data, NPs were placed on a cubic lattice, and then disorder was added. In the dashed and dotted lines, the anisotropy axis direction was changed between each loop. For the other data, the anisotropy axis orientation *and* the NP position was changed between each loop. The dots represent raw data, whereas the other curves represent 200 smoothed points.

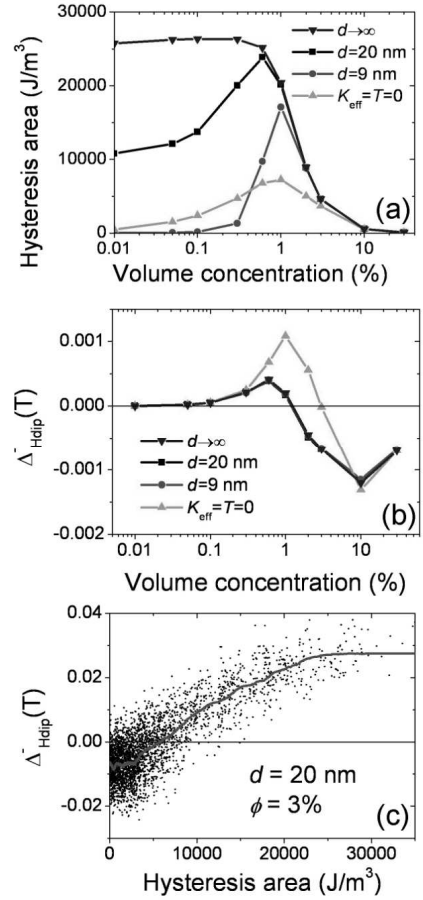

Figure 4 (color online): (a) Hysteresis area of 3000 NP lysosomes as a function of their volume concentration $\phi$. Unless otherwise specified, the parameters were $T = 300$ K, $K_{eff} = 13000$ Jm$^{-3}$, $M_S = 10^6$ Am$^{-1}$, $\mu_0 H_{max} = 40$ mT, $f = 100$ kHz, $v_1^0 = v_2^0 = 10^{10}$ Hz. The hysteresis was composed of 2000 steps and averaged over 50 cycles with a change in anisotropy direction and NP position between each cycle. (▼) $d \to \infty$ ($T = 0$) (■) $d = 20$ nm. (●) $d = 9$ nm. (▲) $K_{eff} = T = 0$. (b) $\Delta^-_{Hdip}$ as a function of the volume concentration. The parameters are identical to the previous graph. (c) Correlation between $\Delta^-_{Hdip}$ and the hysteresis area in a lysosome with $d = 20$ nm and $\phi = 3\%$. Each dot represents a NP. The plain line represents a 200 point average of the data.

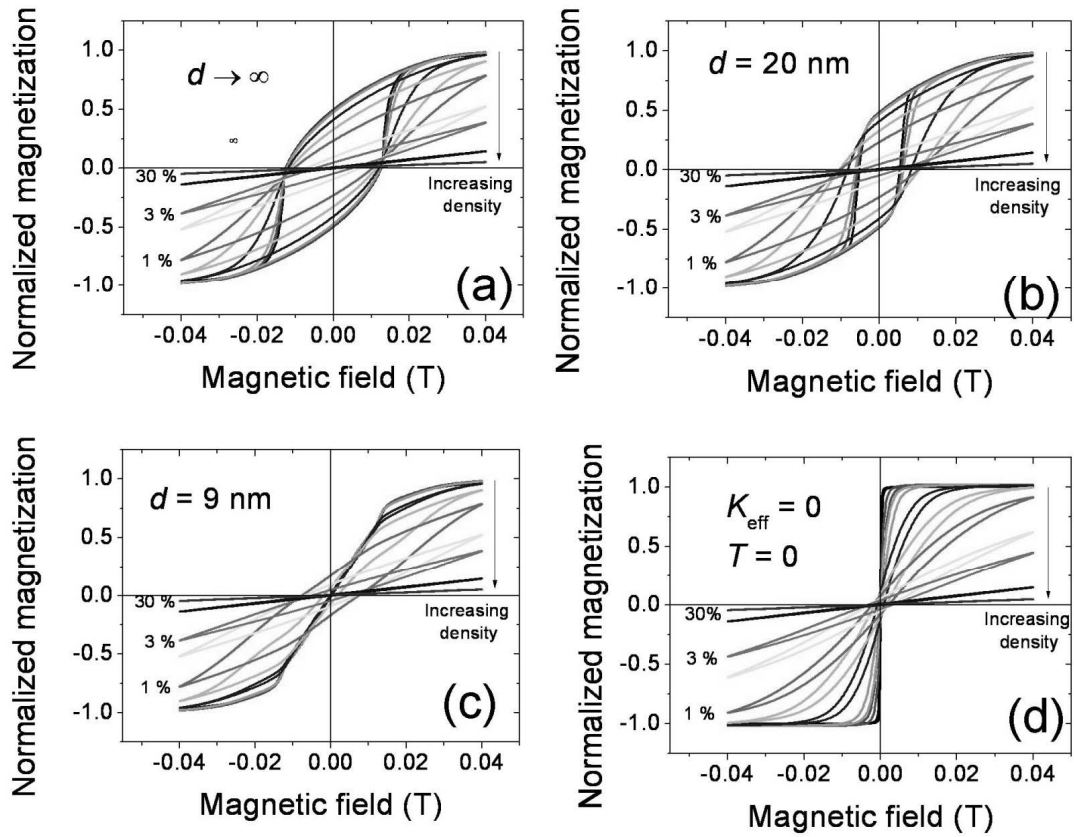

Figure 5 (color online): Hysteresis loops as a function of $\phi$, corresponding to the data of Fig. 4. $\phi$ = 0.01%, 0.05%, 0.1%, 0.3%, 0.6%, 1%, 2%, 3%, 10% and 30%. (a) $d \to \infty$ ($T = 0$). (b) $d = 20$ nm. (c) $d = 9$ nm. (d) $K_{\text{eff}} = T = 0$.

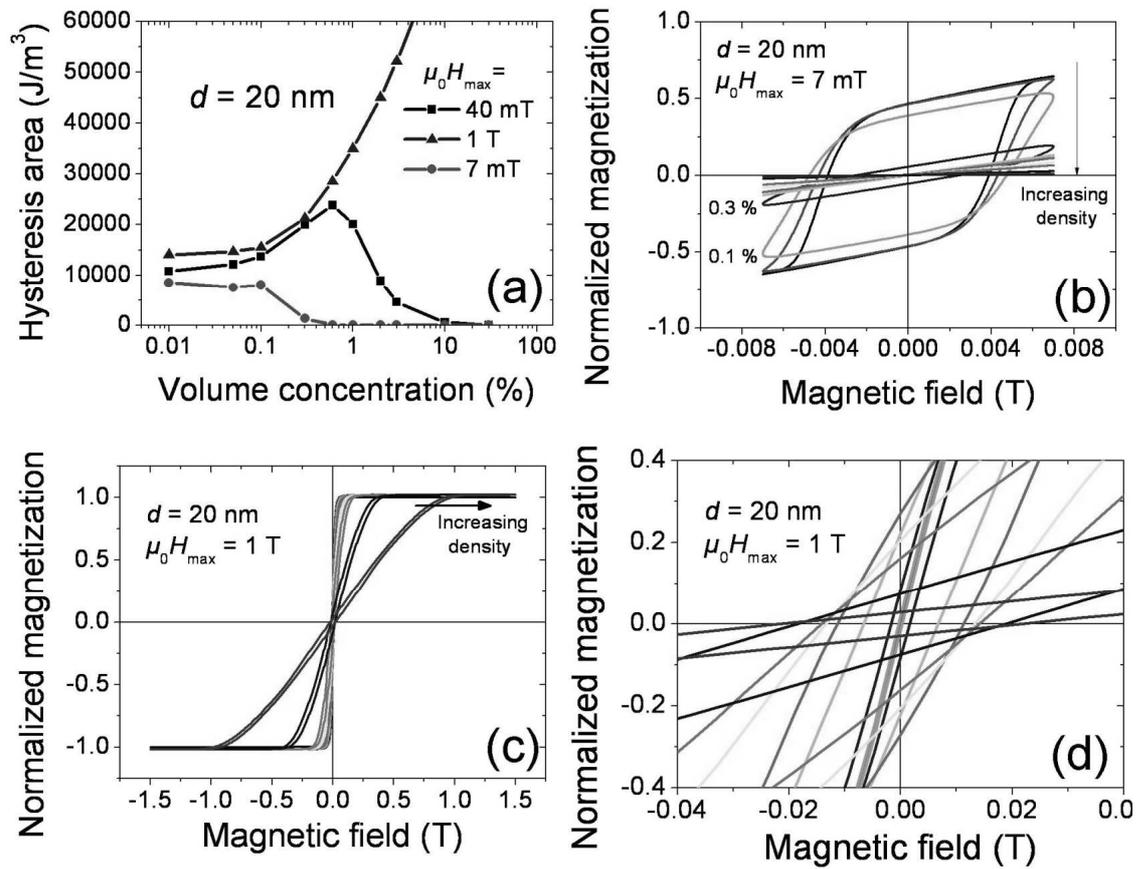

Figure 6 (color online): (a) Evolution of the hysteresis area as a function of $\phi$ for 20 nm NPs. (■) $\mu_0 H_{max} = 40$ mT, (▲) $\mu_0 H_{max} = 1$ T and (●) $\mu_0 H_{max} = 7$ mT. The other simulation parameters were the same as those in Fig. 4. (b), (c), and (d) Corresponding hysteresis loops for (b) $\mu_0 H_{max} = 7$ mT and (c) and (d) $\mu_0 H_{max} = 1$ T. (d) is an enlarged view of (c).

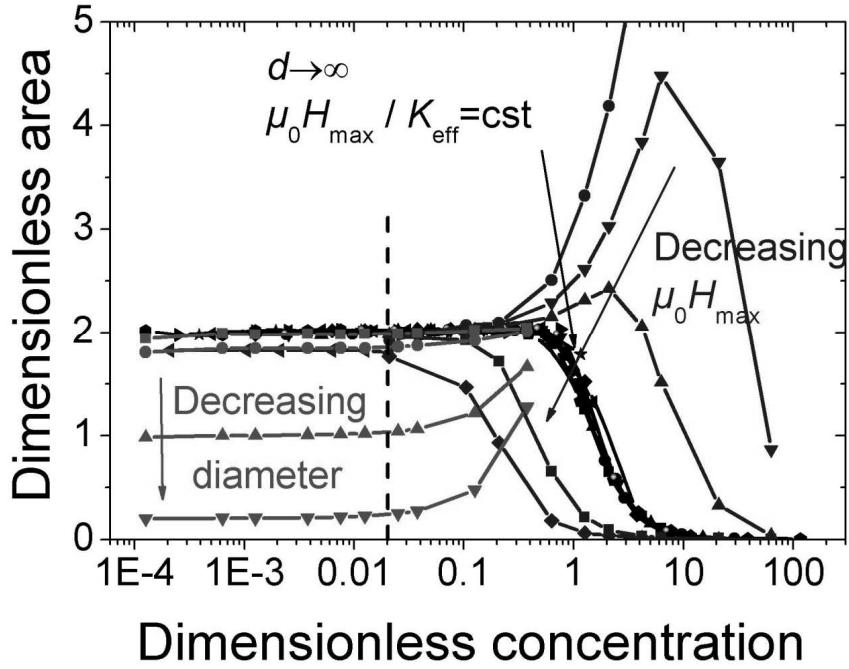

Figure 7 (color online): The dimensionless hysteresis area $\frac{A}{K_{eff}}$ as a function of the normalized concentration $\frac{\mu_0 M_S^2 \phi}{K_{eff}}$. When not specified, the parameters were the same as those for Fig. 4. The vertical, dashed line represents the location where the dimensionless concentration equals 0.02. The black dots and lines represent a series of simulations at $T = 0$ K ($d \to \infty$) with $M_S = 10^6$ Am$^{-1}$ and $K_{eff}$ in the range of $6 \times 10^3$-$1 \times 10^6$ Jm$^{-3}$, as well as a series with $K_{eff} = 13 \times 10^3$ Jm$^{-3}$ and $M_S$ in the range of $2 \times 10^5$-$2 \times 10^6$ Jm$^{-3}$. In these series, the ratio $\frac{\mu_0 H_{max}}{K_{eff}}$ was kept constant. For instance, for $K_{eff}$=6000 J/m$^3$, $\mu_0 H_{max}$=18.4 mT. The blue dots and lines represent a series at $T = 0$ K where $K_{eff} = 6000$ J/m$^3$ and $\mu_0 H_{max}$ varied in the range of 0.008-1 T. The red dots and lines represent a series at $T = 300$ K, $K_{eff} = 10^6$ Jm$^{-3}$ with $d$ in the range of 4-20 nm.

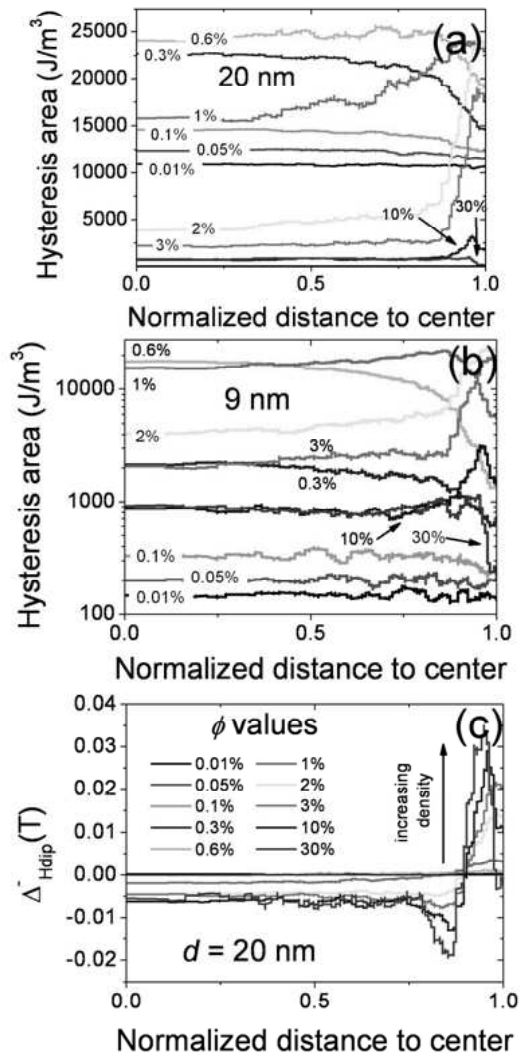

Figure 8 (color online): (a) and (b) Evolution of the hysteresis area as a function of the normalized distance from the center of the lysosome for different $\phi$ values shown on the graphs. The simulation parameters were similar to those from Fig. 4. The curves are the results of 200 smoothed points. (a) $d = 20$ nm. (b) $d = 9$ nm. (c) Evolution of $\Delta^-_{Hdip}$ with the normalized distance from the lysosome center, using the same parameters as in (a).

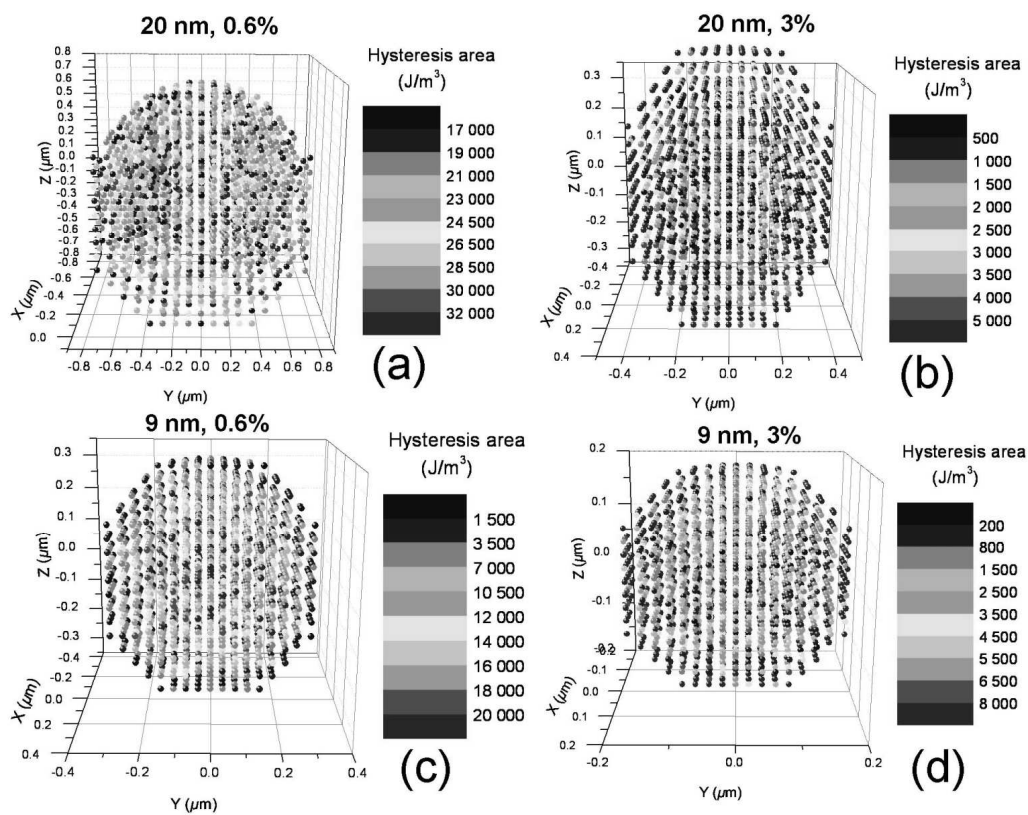

Figure 9 (color online): Heating power of NPs inside a lysosome. The simulation parameters were similar to those from Fig. 4; therefore, there is a link between these two figures. The plotted heating power corresponds to an average over 50 hysteresis loops with a change in the anisotropy axis direction and the exact NP position between each cycle. The NPs are shown positioned on a cubic lattice, which is thus their *average* position. The heating power displayed here is actually that of each NP and is not spatially averaged. The size of the NPs in the figure has been chosen for clarity reasons and does not match their true size. Only half of the lysosome is shown so the reader faces the hemisphere. (a) $d = 20$ nm, $\phi = 0.6\%$. (b) $d = 20$ nm, $\phi = 3\%$. (c) $d = 9$ nm, $\phi = 0.6\%$. (d) $d = 9$ nm, $\phi = 3\%$.

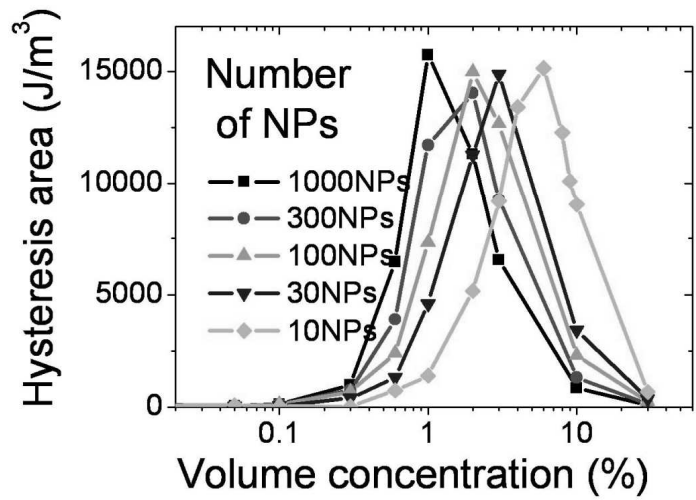

Figure 10 (color online): Hysteresis area as a function of the concentration, plotted for different lysosome sizes. The number of particles inside the lysosomes is shown on the graph. The parameters are the same as those of Fig. 4 with $d = 9$ nm.

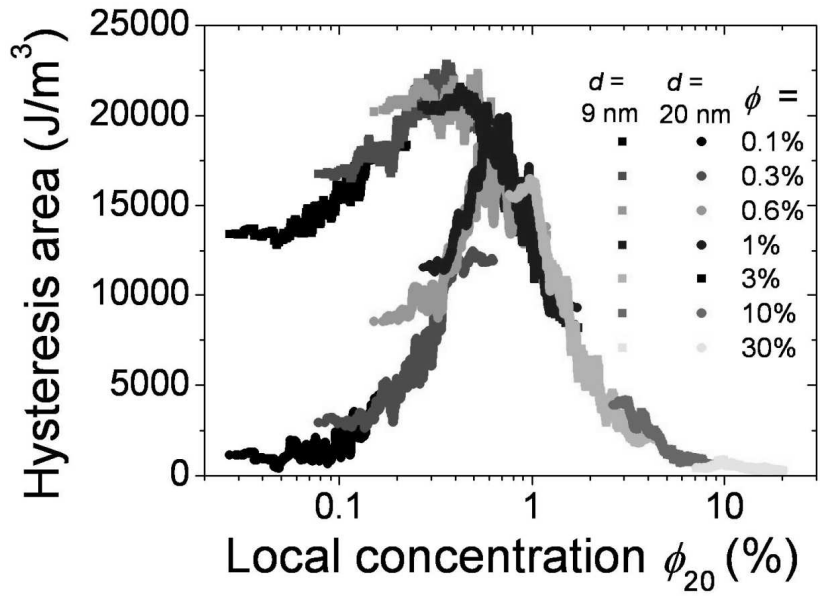

Figure 11 (color online): Hysteresis area as a function of the local volume concentration $\phi_{20}$ calculated using Equ. (8). The simulation parameters were the same as those in Fig. 4 with $K_{\text{eff}} =$ 13000 J/m$^3$ and $T = 300$ K. The results for $d = 9$ nm and $d = 20$ nm are shown. For these results, averaging was only performed on the anisotropy axis direction and not on the position.